\documentclass[aps,twocolumn,pre,showpacs,floatfix]{revtex4}

\usepackage{amsmath,amssymb}
\usepackage{graphics,subfigure}
\usepackage{epsfig}
\usepackage{bm}

\begin{document}

\title{Two-dimensional melting in simple atomic systems 
       : continuous vs. discontinous melting }


\author{Sang Il Lee and Sung Jong Lee}
\affiliation{Department of Physics, University of Suwon, Suwon,
Kyonggi-Do 445-743, Korea}

\begin{abstract}

We investigate the characteristics of two dimensional melting in simple 
atomic systems via isobaric-isothermal ($NPT$) and isochoric-isothermal ($NVT$) 
molecular dynamics simulations with special focus on the effect of the range 
of the potential on the melting. 
We find that the system with interatomic potential of longer range clearly
exhibits a region (in the $PT$ plane) of (thermodynamically) stable hexatic phase. 
On the other hand, the one with shorter range potential exhibits a first-order
melting transition both in $NPT$ and $NVT$ ensembles. Melting of the system with 
intermediate range potential shows a hexatic-like feature near the melting 
transition in $NVT$ ensemble, but it undergoes an unstable hexatic-like phase 
during melting process in $NPT$ ensemble, which implies existence of a weakly 
first order transition. The overall features represent a crossover from a 
continuous melting transition in the cases of longer-ranged potential to a 
discontinuous (first order) one in the systems with shorter and intermediate 
ranged potential. We also calculate the Binder cumulants as well as 
the susceptibility of the bond-orientational order parameter.  

\end{abstract}

\today

\pacs{64.70.Dv, 02.70.Ns, 05.70.Fh, 61.20.Ja} \maketitle

\section{Introduction}

For decades, two dimensional melting\cite{Nelson2002} has been an 
important subject of research in condensed matter physics
both theoretically and experimentally.
One of the most important theoretical frameworks was given by Halperin and
Nelson\cite{Halperin,Nelson1979}, and Young\cite{Young} who proposed 
(building upon the work by Kosterlitz and Thouless\cite{Kosterlitz}) the 
so called KTHNY theory
that the two dimensional melting can occur in two stages of continuous 
defect-mediated transitions with the intermediate hexatic phase 
characterized by quasi-long-range orientational order and short range 
translational order\cite{Strandburg-Review}.

One of the most important questions in the two dimensional melting, which has
not been satisfactorily answered yet, is probably the question of how to 
determine the form of the inter-particle potential that is most favorable for 
the existence of the hexatic phase.
Even though several experimental studies support the existence of hexatic 
phases\cite{Murray87, Kusner94, Marcus96, Zahn99,Pindak81, Brock86, Cheng88, 
Chou1998,Angel2005,Olafsen2005, Reis2006}, computational studies of two-dimensional 
melting of hard-core potential systems\cite{Simulation-Review,Strandburg89,Lee92, Bates2000,
Mak2006} including hard discs or Lennard-Jones (LJ) potentials tend to favor first 
order transition scenarios (though some conflicting results also 
exist)\cite{Strandburg-Review}.
Chui, et al\cite{Chui83-prl, Chui83-prb, Saito1982a,Saito1982b} advanced the possibility 
of first order melting transition through grain boundary formations when the defect core 
energy becomes low enough. Kleinert and Janke argued that the nature of the two-dimensional melting
can change from continuous to first order transition as the magnitude of the so-called 
angular stiffness of the local rotation field\cite{kleinert1988, janke1988, kleinert1989} 
is decreased. From this argument, they contended that the melting of LJ systems
would occur via first order, while, it would be continuous in the case of Wigner crystals
where the particles are interacting via long-range Coulomb potentials.  

\begin{figure}[t]
\includegraphics[angle=0,width=8cm]{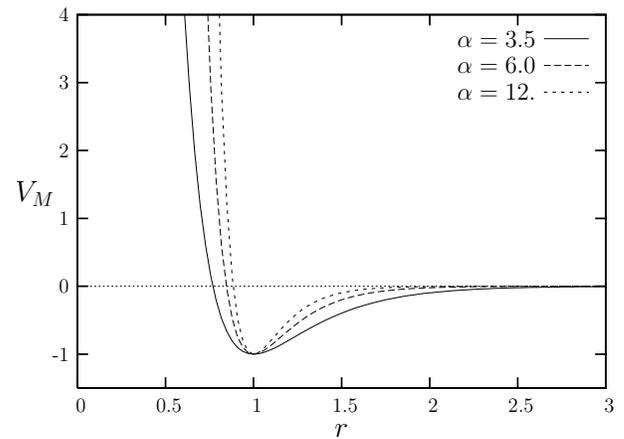}
\caption{Shape of the Morse potential for $\alpha=3.5, 6$ and $12$.} \label{Morse_pot_fig}
\end{figure}

Recently we investigated on the criterion for the existence of the hexatic phase 
by tuning the form of the interparticle potential (which is the Morse potential) 
that could change the size of the range of the dominant interparticle interaction\cite{silee2008}.
The Morse potential can be written as 
\begin{eqnarray}
V_M (r)  & = & \epsilon_0 \left [ e^{ - \alpha (r - \sigma )}  - 1
\right ]^2 - \epsilon_0  \label{Morse_pot}
\end{eqnarray}
where, $r$ is the distance between particles, $\sigma$ the
distance from the origin to the minimum of the potential, and $\epsilon_0 $ is the strength
of the interaction. 
After setting the $\epsilon_0 =1$ and $\sigma =1$, we can vary the value of the single 
parameter $\alpha$ to tune the softness and the range of the potential. 
Smaller value of $\alpha$ corresponds to a softer potential 
with longer range of the attractive part of the potential. On the other hand, as the 
value of $\alpha$ increases, the potential gets stiffer and shorter-ranged (Fig.~\ref{Morse_pot_fig}).

In our previous work, we investigated the trend of hexatic phase formations as 
the range of the potential is varied. Detailed simulation results were presented
especially in the regime of softer potential with $\alpha = 3.0$,
where the melting exhibits a stable region of hexatic phase on th $PT$ plane.

Here in this work, we investigate how the characteristics of melting evolve
as the value of $\alpha$  is varied from $\alpha = 3.5$ to larger values of 
$\alpha = 6$ and $\alpha=12$. See Fig.~\ref{Morse_pot_fig}, for the shape of the 
potentials for different values of $\alpha$.  
We observed that different types of melting are exhibited for an atomic system 
described by the Morse potential. A system with $\alpha=3.5$ exhibits a melting 
transition which is compatible with the KTHNY theory, showing thermodynamically stable 
hexatic phase as well as two stage melting. For simulations with $\alpha=6$, it is 
found that, although some hexatic-like features are revealed in $NVT$ ensemble simulations,
it exhibits a weakly first order transition in $NPT$ ensemble. Systems with a short 
range potential with $\alpha=12$ show a strong first order transition compared to 
the case of $\alpha=6$, clearly showing coexistence of solid and liquid. 
These results appear to be consistent with the arguments of Kleinert\cite{kleinert1989} 
in that the systems with short-ranged potential are correlated with smaller angular
 stiffness and first order melting transition.

\section{Simulation Methods and Results}

In this work, we performed $NPT$ MD simulations using the modified 
Parrinello-Rahman (PR) method\cite{pr1980, Li92} combined
with Nose-Hoover (NH) thermostat\cite{Nose84}. As for the mass of
the particles, for convenience, we put $m=1$ which implies that the
time unit $t_0 \equiv  \sqrt{m \sigma^2 /\epsilon_0}$ also becomes
unity when we set $\sigma = 1$ and $\epsilon_0 =1$. 
The equations of motion were integrated via the
Nordsieck-Gear 5th-order predictor-corrector method with the
integration time step of $\Delta t=0.002$. This guarantees the
conservation of the total Hamiltonian without noticeable drift. In
the simulations we used two empirical parameters, the barostat mass 
$M_{v}=1$ and the thermostat mass $M_{s}=1$. Test simulations
with several other values ($M_{v}=0.1, 1, 10 , M_{s}=0.1, 1, 10$) of
the parameters were also performed with almost the same results.
The number of particles employed ranges from $N=400$ up to $N=10000$.  

In order to investigate the characteristics of the melting transition,
we obtained the isothermal equation of state on the plane of pressure 
vs. density. This was obtained by the NH-MD simulations by decoupling 
the PR (isobaric) part from NH-PR MD equations of motions by taking 
$M_v = \infty$ which reduces the system to $NVT$ condition. 
The pressure was evaluated by means of the virial expression 
for the range of the densities corresponding to the region of transition 
from liquid to solid. For each density, $10^6 \sim 3 \times 10^6$ steps of
integration were carried out for equilibration beginning with a
configuration of triangular lattice and, after equilibration, $10^7$
steps of integration were performed for thermodynamic calculations.

In this case of $NVT$ ensemble we have to fix the shape of the box. 
In order to reduce the finite size effect, we used a rhombic box (with the 
smaller side angle of $60$ degrees) for the shape of the system with periodic 
boundary conditions. However, independent results of ours from square box 
showed no significant difference (from those of rhombic box) with respect 
to the quantities of our interest.

Important criterion for the existence of hexatic phase (and hence continuous
melting transition) would be that the isothermal equation of state for pressure 
vs. density exhibit a monotonic behavior together with a non-monotonic region of 
dip in the slope $dP/d\rho$. On the other hand, a first order melting transition 
would be associated with the existence of a van der Waals type loop in pressure
vs. density curve with unstable and metastable region in $NVT$ ensemble simulations. 


In order to investigate the nature of the possible hexatic phases, one has to
compute the bond-orientational order parameter. The local bond-orientational order 
parameter $\psi_{6}(r)$ at position $r$ is defined as
\begin{equation}
\psi_{6}(r)=\frac{1}{N_{i}}\sum_{j}e^{6i\theta_{ij}(r)}. 
\end{equation}
Here, the sum on particle $j$ is over the $N_{i}$ neighbors of the 
particle $i$ (corresponding to $\vec{r}$ at the center with $\theta_{ij}$
being the angle between the particles $i$ and ${j}$ with respect to
a fixed reference axis.
We regarded the particles within a cutoff radius as the neighbors,
where the cutoff radius is chosen as the first minimum of the pair
correlation function of the system. This method is found to be
efficient and reliable for large scale simulations\cite{Bagchi96}.

Then the global bond-orientational order parameter is defined as   
\begin{equation}
\Psi_{6} = \large | \frac{1}{N}\sum_{r} \psi_{6}(r) \large | 
\end{equation}
where $N$ denotes the total number of particles in the system.
In order to distinguish the bond-orientational order of the different thermodynamic
 phases, we compute the spatial correlation function $G_{6}(r)$ of the 
bond-orientational order parameter, defined as \cite{Halperin}
\begin{equation}
G_{6}(r)=<\psi_{6}(r)\psi_{6}^{\ast}(0)>,
\end{equation}

In the hexatic phase, according to KTHNY theory, the bond-orientational 
correlation function is expected to exhibit an algebraic decay
i.e., $G_{6}(r) \sim r^{-\eta_{6}}$ with the decay exponent $\eta_6 \leq 1/4 $, 
where $\eta_{6}=1/4$ corresponds to the limit of the power-law decay behavior 
in the KTHNY theory. 



In order to further understand the nature of the hexatic phase 
we obtained the histogram distribution\cite{Lee90} of the density order parameter 
for different system sizes for the values of pressure and temperature where 
a melting transition is expected to occur (from other measurements).
We expect that histograms with single peaks would imply that there exists 
continuous melting transition. On the other hand, existence of double peaks 
with increasing peak heights (as the system size increases) would imply 
a first order transition. 




We also investigate the behavior of the linear susceptibility for the global 
bond-orientational order parameter near the melting transition in order to check
the consistency with the result from the isothermal equation of states. 
Specifically, we obtain the size dependence of the susceptibility by calculating
the fluctuation of the bond-orientational order parameter for sub-blocks of the system
with linear sizes $L$, which is defined as\cite{weber95}   
\begin{equation}
\chi_L = L^d \left ( \left < \Psi_{6}^2 \right >_L - \left < \Psi_{6} \right >^{2}_{L} \right ) . 
\end{equation}
where $d=2$ is the spatial dimension. 
In the computation, the system is sub-divided into sub-blocks of linear sizes with
$L = N/ M_b $ where $M_b$ ranges from $10$ to $20$. 

Also, it is useful to calculate the Binder cumulant for the global bond-orientationl 
order parameter for subsystem (linear) size $L$ is defined by 
\begin{equation}
U_{L} = 1- \frac{\left < \Psi_{6}^{4} \right >_L }{ 3 \left < \Psi_{6}^{2} \right >^{2}_{L} }.  
\end{equation}
where the subscript $L$ denotes that the quantities are calculates for subsystem 
sizes of linear size $L$.   


\subsection{The case of a soft and long-ranged Morse potential: $\alpha = 3.5$}

\begin{figure}[t]
\includegraphics[angle=0,width=8cm]{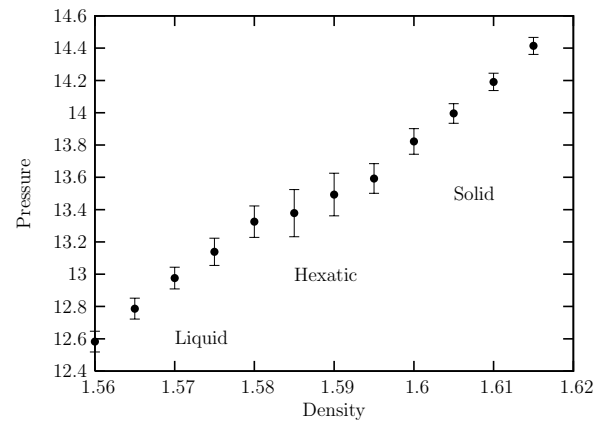}
\caption{ Isothermal equation of state (pressure vs. density) for $\alpha=3.5$ 
at the temperature $T= 0.7$  obtained from $NVT$ ensemble with $N=3600$ 
particles (see the text for details).} \label{pre_den_3.5}
\end{figure}

Here, we first deal with the case of a moderately soft (and longer ranged) 
potential with $\alpha = 3.5$. Figure~\ref{pre_den_3.5} shows the
isothermal equation of state (at $T=0.7$) on the plane of pressure vs.
density. This was obtained by the NH-MD simulations by
decoupling the PR (isobaric) part from NH-PR MD equations of motions
by taking $M_v = \infty$ which reduces to $NVT$ condition. We here define 
the density $\rho $ as $ \rho \equiv N \sigma^2 /V $ where $N$ is the total 
number of particles and $V$ the
total volume (area in two dimensions) of the system. Densities were
chosen from the range $\rho=1.56 \sim 1.62 $, with the density
increment of $\Delta \rho = 0.005$ and the pressure was evaluated by
means of the virial expression (with $k_{B}=1$). This range of the
density corresponds to the region of transition from liquid to
solid. For each density, $10^6 \sim 3 \times 10^6$ steps of
integration were carried out for equilibration beginning with a
configuration of triangular lattice and, after equilibration, $10^7$
steps of integration were performed for thermodynamic calculations.

The number of particles employed was $N=3600$. In order to reduce the
finite size boundary effect, we used a rhombic box (with the smaller
side angle of $60$ degrees) for the shape of the system with
periodic boundary conditions. However, independent results of ours
from square box showed no significant difference (from those of
rhombic box) with respect to the quantities of our interest.

The isothermal curve increases monotonically near the transition
region satisfying the condition of mechanical stability (unlike the
discontinuity of density in a first order transition) that the
isothermal compressibility should be positive 
$ K_T = (1/\rho) (\partial \rho / \partial P)_T  > 0$. 
We may identify the boundary of stable hexatic
phase as the values of the density where an abrupt change in the
isothermal compressibility occurs. In this way, we estimate the
density of solid-hexatic transition as $\rho_{s-h} \simeq 1.6$.

Although the change in isothermal compressibility is less
conspicuous near the hexatic-liquid boundary, we see that, near the
density $1.58\leq \rho \leq 1.585$, there exists a crossover in the
slope of the isothermal compressibility. Below, we give an
estimation of the density of hexatic-liquid transition by applying a
theoretical expectation from KTHNY theory on the decay exponent of
the spatial correlation of the orientational order parameter (see
below).

The fact that the pressure within the hexatic phase is monotonically
increasing as the density increases (with the resulting isothermal
compressibility kept positive) appears to be a very compelling
evidence for a stable hexatic phase in thermal equilibrium.

In order to distinguish the orientational order of the phases, we
have computed the  bond-orientational correlation function $G_{6}(r)$
defined above.

\begin{figure}[t]
\includegraphics[angle=0,width=8cm]{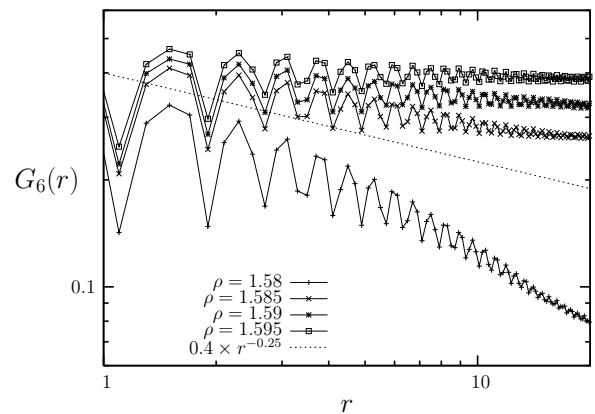}
\caption{Spatial correlation functions (for $\alpha = 3.5$) of the 
bond-orientational order parameter for different densities at $T=0.7$.
The dashed line indicates a power law decay with the decay exponent of
$1/4$.} \label{r_ori_3.5}
\end{figure}

Figure.~\ref{r_ori_3.5} is the bond-orientational correlation function for
the range of the density ($1.56\leq \rho \leq 1.615$). We find that,
for the density range of $1.585 < \rho \leq 1.595$, the averaged
correlation functions exhibit algebraic decays with the decay
exponent $\eta < 1/4 $  while, at lower densities, they exhibit 
decays with faster than power law behavior in the long distance limit. 
At $\rho=1.585$, the orientational correlation exhibit a slope of
approximately $\eta=0.25$. Here, the crossover from a power law decay 
to exponential takes places with exponent approximately equal to $1/4$, 
and also this value of the crossover density agrees almost precisely with 
the density region exhibiting an abrupt change of the slope i.e., 
of the compressiblity in the equation of state (Figure~\ref{pre_den_3.5}).

\begin{figure}[t]
\includegraphics[angle=0,width=8cm]{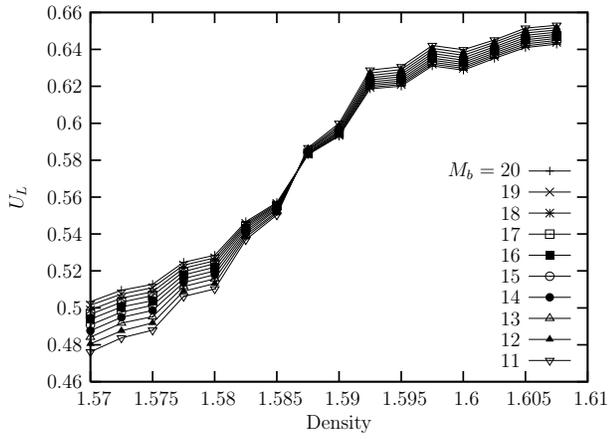}
\caption{Fourth order Binder cumulant of the bond-orientational order parameter 
vs. density obtained by sub-block method at $T=0.7$ ($\alpha=3.5$) obtained from
$NVT$ ensemble. 
The total number of particles is $N=10000$ ($100 \times 100$ and the sub-blocks have 
dimensions of $L\times L = (N/M_b)\times (N/M_b)$ with $M_b$ ranging 
from $11$ to $20$.} \label{binder_3.5}
\end{figure}

Now, the fourth order Binder cumulant for the global bond-orientationl 
order parameter for sub-block systems of (linear) size $L$ is shown in
Fig.~\ref{binder_3.5}. We can see that the density at the crossing point   
is around $\rho \simeq 1.585$. This is compatible with the boundary 
density ($1.585$) between the liquid and the hexatic phase which was shown 
above in isothermal curve for $NVT$ ensemble, and also compatible with the 
the boundary density between liquid and hexatic-like phase in terms of the 
power law decay of the bond-orientational order near $\rho =1.585$

It may be expected theoretically that the binder cumulants of local orientational
order in hexatic phase collapse to a line because of the critical 
charateristic of the phase. However, we may not consider non-collapse of 
the Binder cumulants to a line as an evidence for non-existence of hexatic 
phases. This is because, even for the case of XY model,  
complete collapse was not found in the region where orientational order decay 
algebraically, but rather it exhibits a crossing point at the transition 
temperature \cite{oliveira}. 

\begin{figure}[t]
\includegraphics[angle=0,width=8cm]{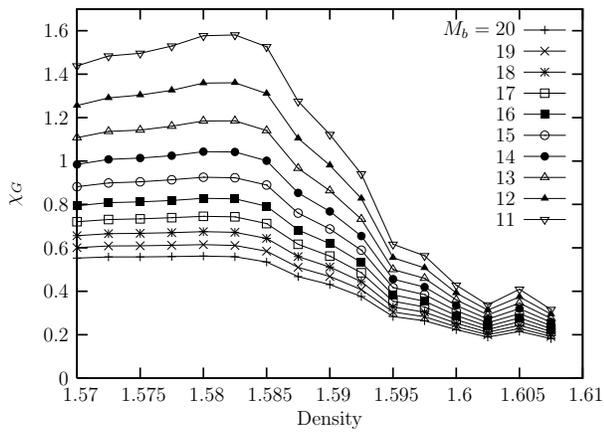}
\caption{Orientational susceptibility vs. density, obtained from the 
fluctuation of the bond-orientational order parameter using sub-block method
with the same system as in Fig.~\ref{binder_3.5} at 
$T=0.7$ ($\alpha=3.5$).} \label{suscep_3.5}
\end{figure}

Now, we turn to the linear susceptibility for the global bond-orientational order 
parameters near the melting transition. 
Shown in Fig.~\ref{suscep_3.5} is the sub-block susceptibility obtained from the 
fluctuation of the bond-orientational order parameter for sub-blocks of the system
with linear sizes $L$ with $L = N/ M_b $ where $M_b$ ranges from $11$ to $20$. 
We see that the suceptibility shows a broad peak region near the density 
$ 1.580 < \rho < 1.585$ which borders the liquid-hexatic phase boundary region. 
We also see that the suceptibility exhibits broader shape (showing weaker dependence
on the density) in the liquid region compared with other cases that will be 
shown below.


Figure~\ref{p3.5_conf_9} shows a snapshot of the particle configuration 
for density $\rho=1.585$ within the hexatic phase region but close to 
the transition (to liquid phase) at the temperature $T=0.7$, which shows free 
dislocations (i.e., bounded pair disclinations). 
This shows rather clearly the fundamental role of defects leading to the
power law decay of orientational correlations.

\begin{figure}[t]
\includegraphics[angle=0,width=8cm]{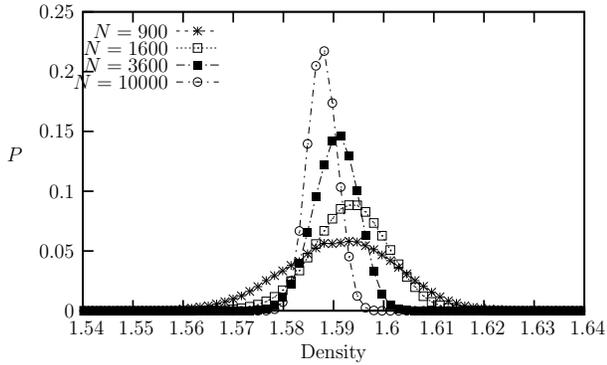}
\caption{Histogram distributions (for $\alpha = 3.5$) of the particle density 
from $NPT$ ensemble simulations at $P=13.5, T=0.7$ for system sizes 
$N=900$, $1600$, $3600$, and $10000$, respectively.} 
\label{den_dist_Morse_3.5}
\end{figure}

In order to further understand the nature of the hexatic phase 
we obtained the histogram distribution\cite{Lee90} of the density order parameter 
for five different system sizes ($N=900, 1600, 3600, 10000$) 
under constant external pressure and temperature of $T=0.7$, and $P=13.5$, where 
a hexatic phase is expected to occur from our measurement of the
orientational correlations. 
In Fig.~\ref{den_dist_Morse_3.5} we see that all the
histograms exhibit single peaks, which implies that there exist a unique phase 
with minimum free energy. It is also observed that, as the number of 
particles increases, the peak height becomes higher and, at the same time,
the width of the peak decreases. 
Also the position of the peak tends to shift to the lower density 
(within the hexatic regime) probably due to the development of long range fluctuation. 
This indicates that this region corresponds to a single phase region 
(unlike solid-liquid mixture) consistent with the absence of van der Waals loop in 
the pressure vs. density curve.

All of these observations lead us to the conclusion that there exist a 
thermodynamcally stable hexatic phase consistent with the KTHNY melting scenario
for the case of $\alpha = 3.5$.

\subsection{The case of an intermediate-ranged potential: $\alpha = 6$ }

\begin{figure}[t]
\includegraphics[angle=0,width=8cm]{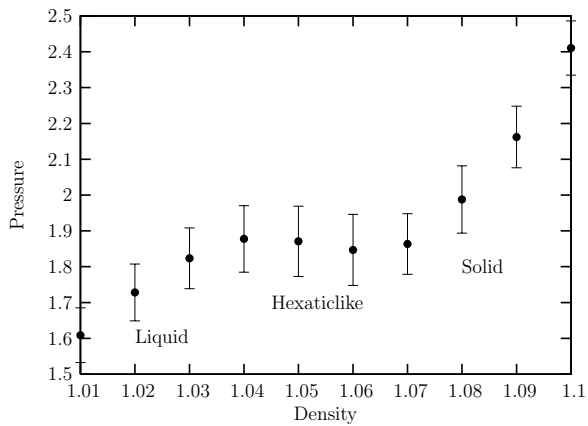}
\caption{Isothermal equation of state (pressure vs. density) obtained from
$NVT$ ensemble with $N=3600$ (for $\alpha=6$) at temperature $T= 0.57$ 
which shows a weak van der Waals-like loop.} \label{pre_den_6}
\end{figure}

\begin{figure}[t]
\includegraphics[angle=0,width=8cm]{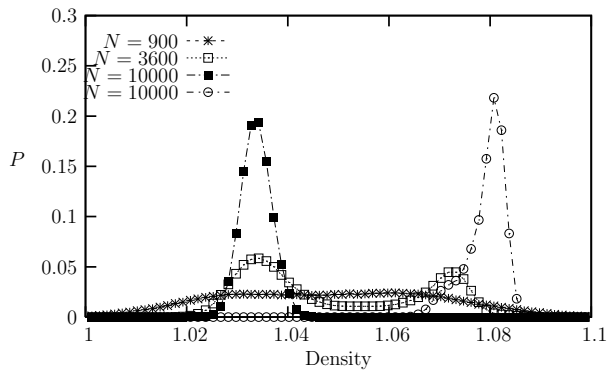}
\caption{Histogram distributions of the particle density (for $\alpha=6$) from 
isobaric-isothermal ensemble simulations at $T=0.57$, $P=1.85$ for system sizes 
$N=900$, $3600$, and $10000$. Note that we obtain two separate histograms for the 
system size $N=10000$ with ordered initial states (higher density) and with disordered
initial states (lower density), respectively. This is because the system of $N=10000$ 
does not exhibit transitions between liquid-like states to solid-like states during 
$10^8$ MD steps with $\Delta T=0.002$.} \label{den_dist_Morse6}
\end{figure}

\begin{figure}[t]
\includegraphics[angle=0,width=8cm]{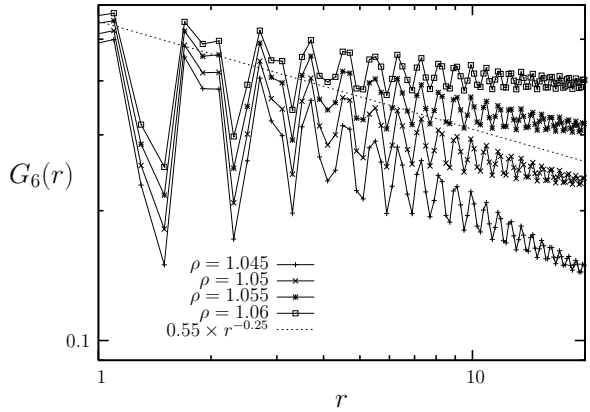}
\caption{Spatial correlation functions of the bond-orientational order parameter for 
different densities (for $\alpha = 6$).
The dashed line indicates a power law decay with the decay exponent of $1/4$.
} \label{r_ori6}
\end{figure}

Next, we examine an intermediate-range potential with $\alpha=6$,
which corresponds approximately to the famous LJ potential\cite{tclim2003}. 
In Fig.~\ref{pre_den_6}, the equation of state exhibits a weak van der Waals-like loop
in the pressure vs. density, which indicates a first order transition. 
The unstable region ranges from $ \rho \simeq 1.04 $ to $ \rho \simeq 1.065$.  
This is confirmed more rigorously by the histogram distributions of the density in 
$NPT$ ensemble simulations for different system sizes. 
In Fig.~\ref{den_dist_Morse6}, the histogram distributions of the density for systems
with $N=900$, $3600$, and $10000$ are shown from $NVT$ ensemble for $T=0.57$ and $P=1.85$.
For the cases of $N=900$ and $3600$, we observe transitions between two peaks (through 
the valley of finite height between the peaks) with the resulting double peaks in the 
histograms. For the $N=10000$ systems, 
however, we can no longer observe crossing between the the coexisting phases. Instead, 
we could observe two different (separate) histograms that are determined by the 
initial states, depending whether the initial state is in the ordered solid 
phase or in the disordered liquid phase.  
Evidently, the free energy barrier increases with increasing system size, which
indicates clearly that the transition is of first order\cite{Lee90}. 
Nevertheless, the system configurations in the coexistence region resembles those of the
hexatic phases (Fig.~\ref{p6_conf_11}), showing algebraically decaying orientational
order (Fig~\ref{r_ori6}). We see that the boundary between liquid and hexatic-like 
phase in terms of the orientational order is located around  $ \rho = 1.05 \sim  1.055$.

Furthermore, we also observe a hexatic-like feature in 
the $NPT$ ensemble, where the system goes through temporarily a hexatic-like phase  
before transiting into the other phase. This kind of characteristics in $NPT$ ensemble 
seems to have been already reported as `metastable hexatic phase' in LJ system by 
Chen et al\cite{Chen95, Somer97, Somer98}. 
Although they argued that large system size ($N \simeq 40000$) is necessary to observe 
this kind of features, we could observe such metastable hexatic phases even for systems with
smaller sizes of $N=1600$. We think that this hexatic-like feature in a first order transition 
can be attributed to weakly first order nature of the transition. 
As shown below, the relative free energy barrier in this case is considerably lower than that
in the case of shorter-ranged potential of $\alpha=12$, from which we presume that 
the metastable or unstable hexatic-like phase comes from defect proliferation by 
thermal fuctuations, but not from some mechanism leading to a true phase transition.

\begin{figure}[t]
\includegraphics[angle=0,width=8cm]{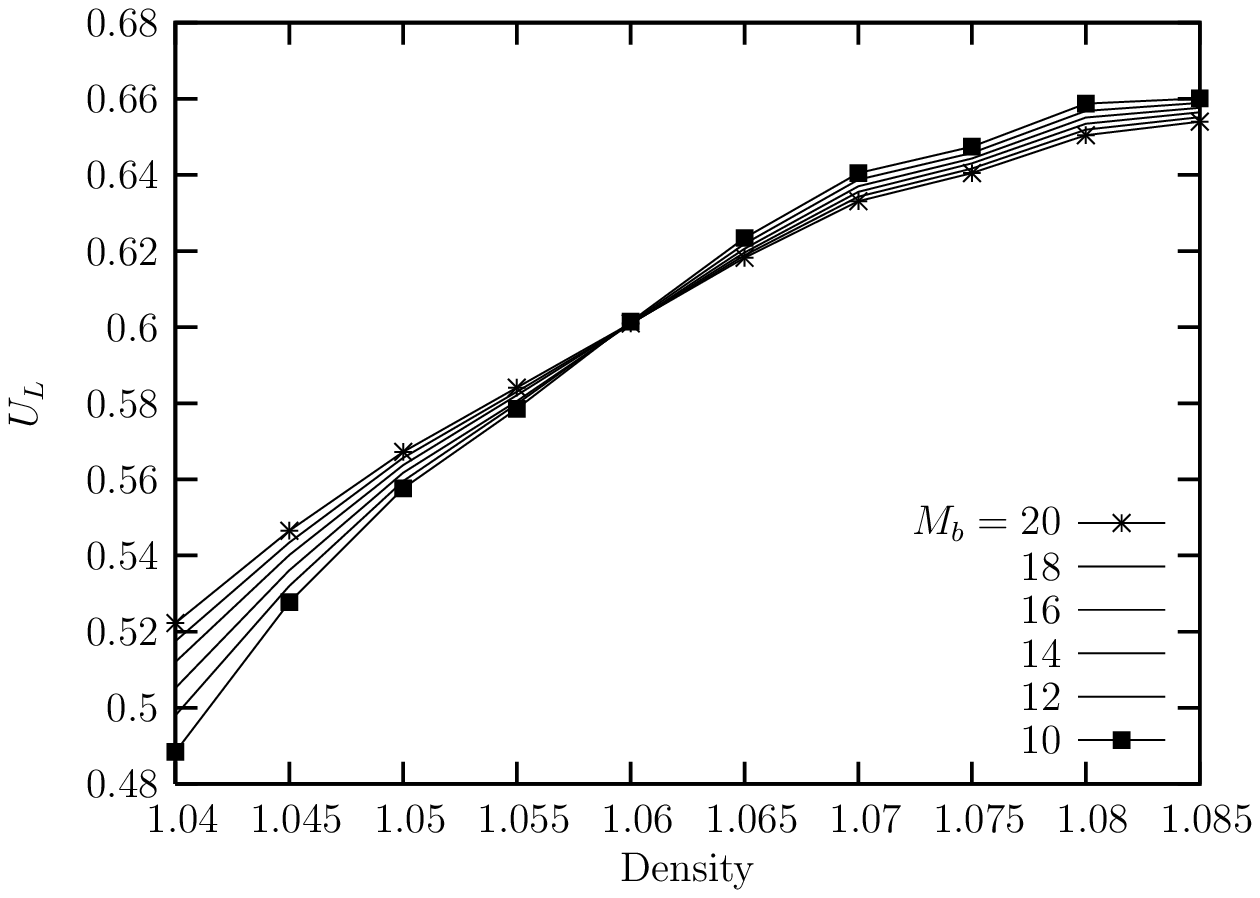}
\caption{Fourth order Binder cumulant of the bond-orientational order parameter
vs. density obtained by sub-block method at $T=0.57$ (for $\alpha=6$)
The total number of particles is $N=10000$ ($100 \times 100$ and the sub-blocks have 
dimensions of $L\times L = (N/M_b)\times (N/M_b)$ with $M_b$ ranging 
from $10$ to $20$.} \label{binder_6.0}
\end{figure}

Now, the fourth order Binder cumulant for the global bond-orientationl 
order parameter for sub-block systems of (linear) size $L$ is shown in
Fig.~\ref{binder_6.0}. We can see that the density at the crossing point   
is located near $\rho \sim 1.06$. This corresponds to a point 
in the middle of the unstable part of the van der Waals curve in the
isothermal equation of state. 

\begin{figure}[t]
\includegraphics[angle=0,width=8cm]{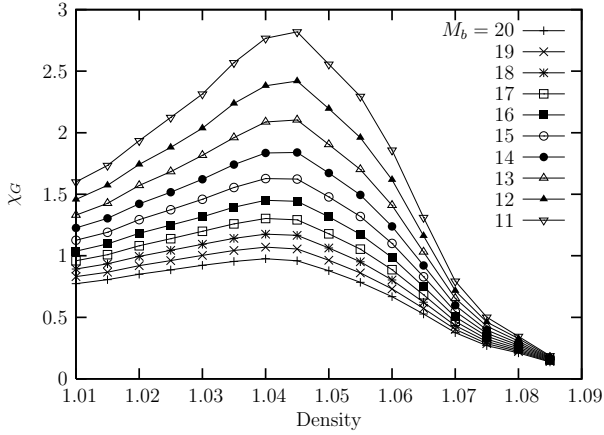}
\caption{Orientational susceptibility obtained from the fluctuation of the bond-orientational 
order parameter using sub-block method with the same system as in Fig.~\ref{binder_6.0} at 
$T=0.57$ (for $\alpha=6$). } \label{suscep_6.0}
\end{figure}

Next, we deal with the linear susceptibility for the bond-orientational order 
parameters. Shown in Fig.~\ref{suscep_6.0} is the sub-block susceptibility obtained 
from the fluctuation of the bond-orientational order parameter for sub-blocks of 
the system with linear sizes $L$ with $L = N/ M_b $ where $M_b$ ranges from $10$ to $20$. 
We see that the suceptibility shows a sharper peak (as compared with the case of
$\alpha =3.5$) at the density $\rho \simeq 1.045$ which is a little bit below 
the density ($\rho = 1.05 \sim 1.055$) where the orientational correlation 
exhibits a spatial decay exponent of $0.25$. 
This might be interpreted as a small evidence that the nature of the melting transition 
of this system is inconsistent with the expectation of the KTHNY theory.


\subsection{The case of a short-ranged potential: $\alpha = 12$}

\begin{figure}[t]
\includegraphics[angle=0,width=8cm]{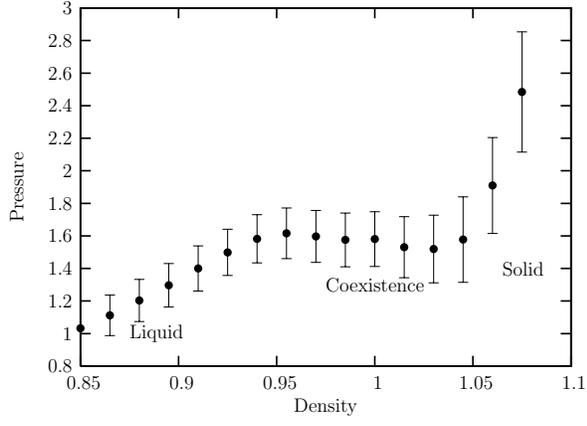}
\caption{Isothermal equation of state (pressure vs. density) obtained from
$NVT$ ensemble with $N=3600$ (for $\alpha=12$) at temperature $T= 0.57$ 
exhibiting a clear van der Waals-like loop.} \label{pre_dist_12}
\end{figure}

 Finally, we investigate the case of a short-ranged potential with $\alpha=12$.
Figure.~\ref{pre_dist_12} shows the equation of state in the density region
of $0.85 \leq\rho\leq 1.075$ obtained from $NVT$ ensemble simulations at $T=0.57$.
We can see that the equation of state exhibits a van der Waals-like region in the
density with the unstable region ranging from $ \rho \simeq 0.96$ to $ \rho \simeq 1.04$
clearly indicating a first order melting transition.

\begin{figure}[t]
\includegraphics[angle=0,width=8cm]{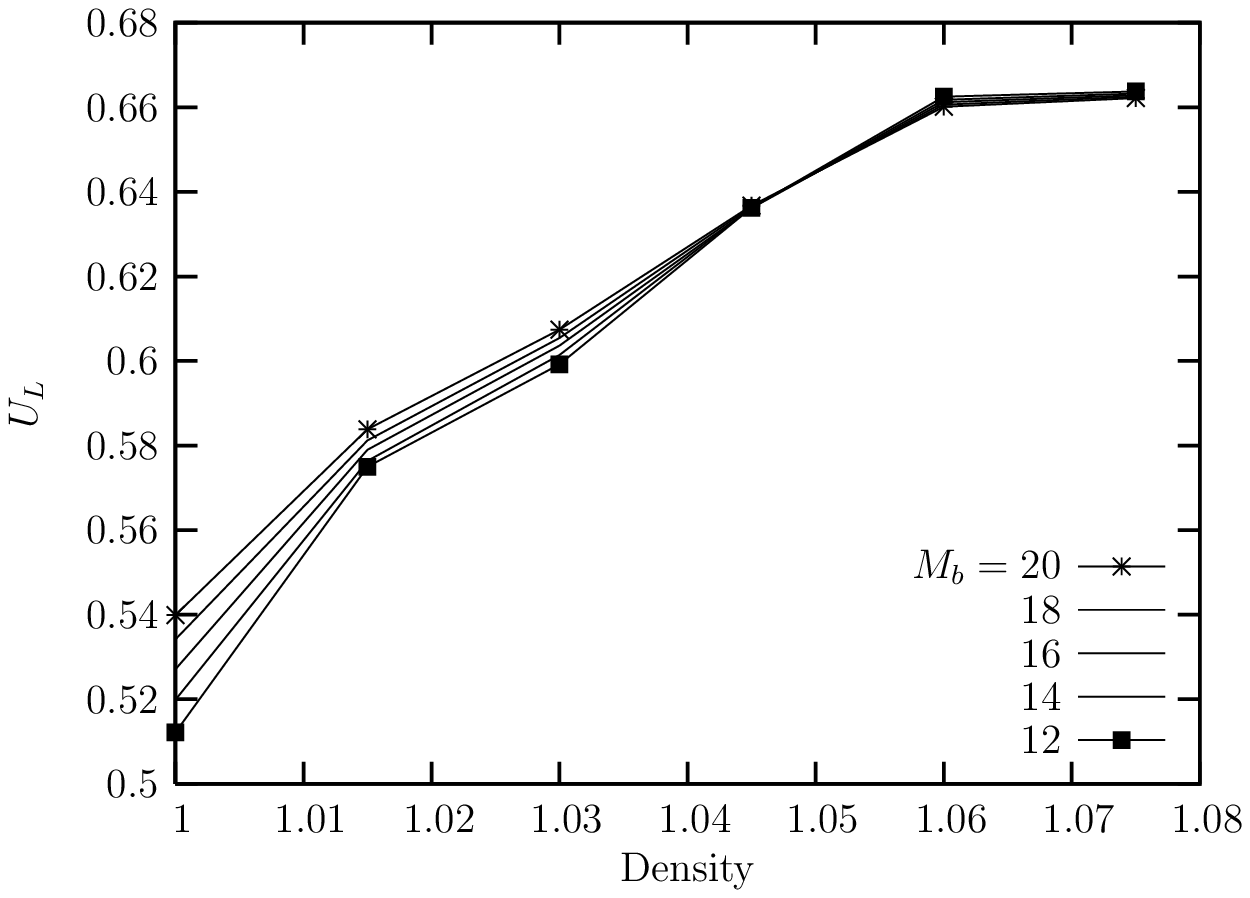}
\caption{Fourth order Binder cumulant of the bond-orientational order parameter
vs. density obtained by sub-block method at $T=0.57$ (for $\alpha=12$)
The total number of particles is $N=10000$ ($100 \times 100$ and the sub-blocks have 
dimensions of $L\times L = (N/M_b)\times (N/M_b)$ with $M_b$ ranging 
from $12$ to $20$.} \label{binder_12.0}
\end{figure}

The fourth order Binder cumulant for the global bond-orientationl 
order parameter for sub-block system size $L$ is shown in
Fig.~\ref{binder_12.0}. We can see that the density at the crossing point   
is located near $\rho \sim 1.045$ which is located near the lower density limit of the
metastable solid (spinodal) as shown in the $NVT$ isothermal equation of state.

\begin{figure}[t]
\includegraphics[angle=0,width=8cm]{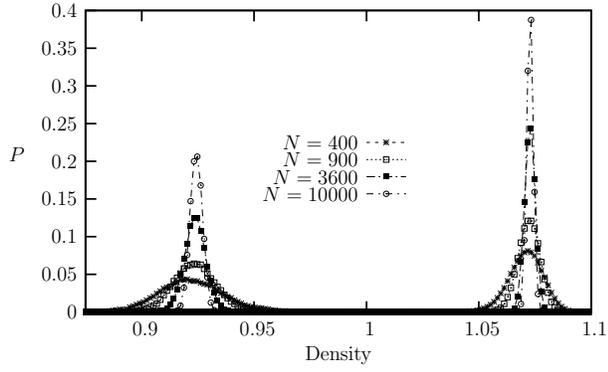}
\caption{Histogram distributions of the particle density (for $\alpha=12$) from 
isobaric-isothermal ensemble simulations at $T=0.57$, $P=1.825$ for system sizes 
$N=400$, $900$, $3600$, and $10000$. Note that none of the systems 
exhibit any tunneling transitions between liquid-like states to solid-like states during
$10^8$ MD steps of simulations with $\Delta T=0.002$.
Therefore, we obtained two separate histograms for each of the system sizes with 
ordered initial states (higher density) and with disordered initial states (lower density), 
respectively.} \label{den_dist_12}
\end{figure}

First order nature of the melting transition is confirmed further with the double peak 
nature of the histogram distributions of the density from $NPT$ ensemble simulations
for system sizes of $N=400, 900, 3600, 10000$ as shown in Fig.~\ref{den_dist_12}. 
In this case, the free energy barrier is presumably much higher than the case of 
$\alpha=6$, and none of the systems with $\alpha =12$ exhibit any tunneling transitions 
between liquid-like states to solid-like states during $10^8$ MD steps of simulations with 
$\Delta T=0.002$. 
Therefore, double peak histogram distributions for each of the system sizes are actually
obtained by combining two separate histograms, one with ordered initial states (higher density) 
and the other with disordered initial states (lower density), respectively. 

Also, typical system configuration for $\alpha =12$ is shown in Fig.~\ref{p12_conf_10}
for $\rho =1.0$ and $T=0.57$ corresponding to the coexisting region, where we can see 
that hexatic-like feature disappears, and that liquid phase region consisting of defect 
clusters coexists with 
solid region. Also in $NPT$ ensemble simulations of the melting process, we can no longer 
observe metastable or unstable hexatic phase, but observe a discontinuous abrupt change in 
density. From these observations we thus conclude that first order nature of the melting 
gets stronger as the potential range decreases.

\begin{figure}[t]
\includegraphics[angle=0,width=8cm]{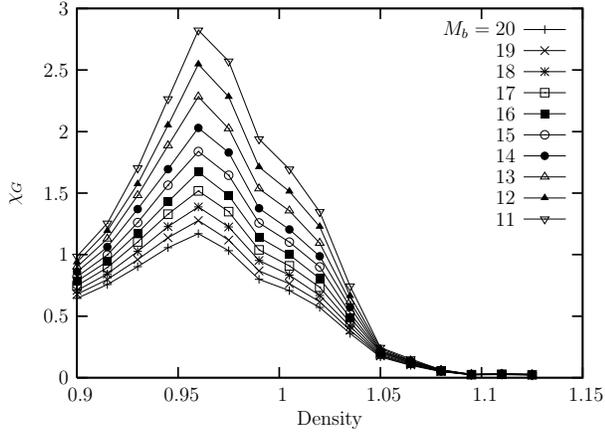}
\caption{Orientational susceptibility obtained from the fluctuation of the bond-orientational 
order parameter using sub-block method at $T=0.57$ (for $\alpha=12$).
The total number of particles is $N=10000$ ($100 \times 100$ and the sub-blocks have 
dimensions of $L\times L = (N/M_b)\times (N/M_b)$ with $M_b$ ranging 
from $11$ to $20$.} \label{suscep_12}
\end{figure}

Next, we look into the linear susceptibility for the bond-orientational order 
parameters. Shown in Fig.~\ref{suscep_12} is the sub-block susceptibility obtained 
from the fluctuation of the bond-orientational order parameter for sub-blocks of 
the system with linear sizes $L$ with $L = N/ M_b $ where $M_b$ ranges from $12$ to $20$. 
We see that the suceptibility shows a peak at the density $\rho \simeq 0.96$ 
which is located near the limit of the metastable liquid (spinodal) as shown 
in the curve of the isothermal equation of state from $NVT$ ensemble.


\section{Summary and discussions}

In conclusion, we have reported on some details of two dimensional melting 
in systems of particles interacting via Morse potential when the range of the 
potential is varied.
We showed that the melting of system with longer-ranged potential ($\alpha=3.5$) 
clearly exhibits features of melting consistent with KTHNY theory exhibiting stable
hexatic phases. As the range of the potential decreases, however, we observe a 
crossover in the transition nature to a first order transtion. In the case of $\alpha=6$
where the range of the potential is intermediate, the system exhibits a weakly first 
order melting transition with some unstable hexatic-like phase during melting process 
in $NPT$ simulations. In the case of $\alpha=12$ where the range of the potential is 
shorter, we observe a stronger first order melting.
It appears that the crossover from continuous to first order melting transition in 
this system is related to the decrease of the so-called angular stiffness
of the rotation field\cite{kleinert1988, janke1988, kleinert1989}. It would be interesting
to carry out a detailed calculation of the angular stiffness in our model system
as the value of $\alpha$ is varied. It would be also interesting to find a possible 
connection to the change from continuous to first order transition in two dimensional 
$XY$ model when the shape of the $XY$ potential gets 
sharpened\cite{domany1984,himber_1984a,himber_1984b}.


\newpage

\medskip

\begin{figure}[t]
\includegraphics[angle=0,width=8cm]{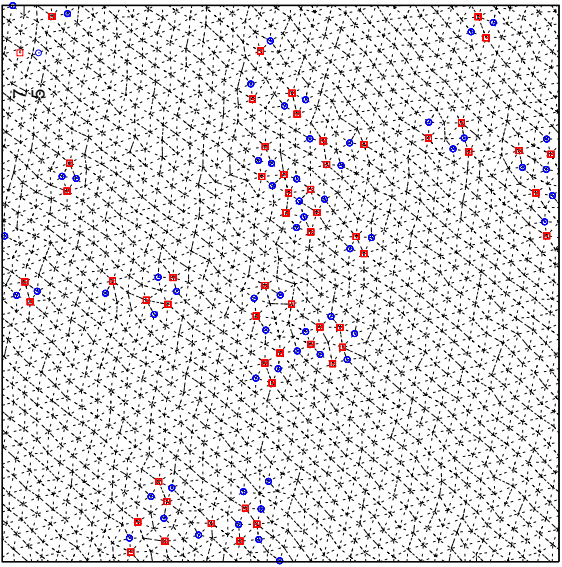}
\caption{(Color online) A snapshot of the configuration of particles at density 
$\rho=1.585$ for $\alpha = 3.5$ ($T=0.7$) represented with Delaunay triangulation. 
Blue open circles and red open squares denote the defect sites of particles with 
five nearest neighbors and seven nearest neighbors, 
respectively.} \label{p3.5_conf_9}
\end{figure}



\begin{figure}[t]
\includegraphics[angle=0,width=8cm]{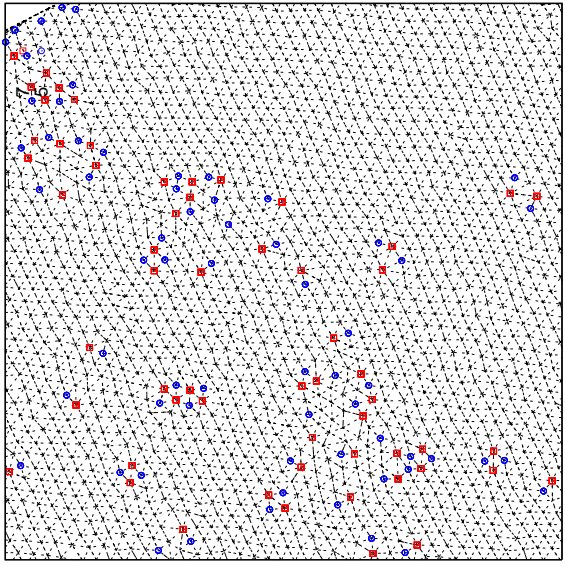}
\caption{(Color online) A snapshot of the configuration of particles at density
$\rho= 1.06$ for $\alpha=6$ ($T=0.57$) represented with Delaunay triangulation.
The defects are indicated with the same method as in Fig.~\ref{p3.5_conf_9}.
} \label{p6_conf_11}
\end{figure}

\begin{figure}[t]
\includegraphics[angle=0,width=8cm]{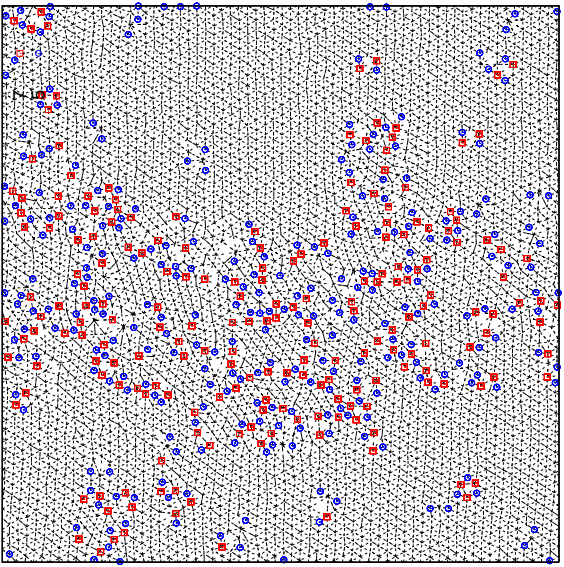}
\caption{(Color online) A snapshot of the system configuration 
at density $\rho= 1.0$ for $\alpha=12$ ($T=0.57$) represented with Delaunay triangulation.
The defects are indicated with the same method as in Fig.~\ref{p3.5_conf_9}.
We can clearly see coexistence of liquid-like region and solid-like region.
} \label{p12_conf_10}
\end{figure}

\end{document}